\documentclass[dvips,preprint]{imsart}

\usepackage{amsthm,amsmath}
\usepackage{amsbsy}

\usepackage{algorithm}
\usepackage[noend]{algpseudocode}

\usepackage{epsf}
\usepackage{epsfig}

\startlocaldefs

\newcommand{\bfx}{\boldsymbol{x}}
\newcommand{\bfz}{\boldsymbol{z}}
\newcommand{\bfone}{\boldsymbol{1}}
\newcommand{\bftht}{\boldsymbol{\theta}}

\newcommand{\bfw}{\boldsymbol{w}}
\newcommand{\bfW}{\boldsymbol{W}}
\newcommand{\bfn}{\boldsymbol{n}}
\newcommand{\bfy}{\boldsymbol{y}}
\def\BState{\State\hskip-\ALG@thistlm}
\endlocaldefs

\begin{document}

\begin{frontmatter}

\title{Speeding up bootstrap computations: a vectorized implementation for statistics based on sample moments}
\runtitle{Vectorized bootstrap for statistics based on sample moments}


\author{\fnms{Elias} \snm{Chaibub Neto}\ead[label=e1]{elias.chaibub.neto@sagebase.org, Sage Bionetworks}}
\address{\printead{e1}}


\runauthor{Chaibub Neto E.}

\begin{abstract}
In this note we propose a vectorized implementation of the non-parametric bootstrap for statistics based on sample moments. Basically, we adopt the multinomial sampling formulation of the non-parametric bootstrap, and compute bootstrap replications of sample moment statistics by simply weighting the observed data according to multinomial counts, instead of evaluating the statistic on a re-sampled version of the observed data. Using this formulation we can generate a matrix of bootstrap weights and compute the entire vector of bootstrap replications with a few matrix multiplications. Vectorization is particularly important for matrix-oriented programming languages such as R, where matrix/vector calculations tend to be faster than scalar operations implemented in a loop. We illustrate the gain in computational speed achieved by the vectorized implementation in real and simulated data sets, when bootstrapping Pearson's sample correlation coefficient.
\end{abstract}





\end{frontmatter}

\section{Introduction}

In this note we propose a vectorized implementation of the non-parametric bootstrap for statistics based on sample moments. Our approach is based on the multinomial sampling formulation of the non-parametric bootstrap. This formulation is described in the next section, but, in essence, follows from the fact that the bootstrap distribution of any sample moment statistic (generated by sampling $N$ data points with replacement from a population of $N$ values, and evaluating the statistic on the re-sampled version of the observed data) can also be generated by weighting the observed data according to multinomial category counts sampled from a multinomial distribution defined over $N$ categories (i.e., data points) and assigning probability $1/N$ to each one of them.

The practical advantage of this multinomial sampling formulation is that, once we generate a matrix of bootstrap weights, we can compute the entire vector of bootstrap replications using a few matrix multiplication operations. The usual re-sampling formulation, on the other hand, is not amenable to such vectorization of computations, since for each bootstrap replication one needs to generate a re-sampled version of the data. Vectorization is particularly important for matrix-oriented programming languages such as R\cite{Rcoreteam2014} and Matlab\cite{matlab}, where matrix/vector computations tend to be faster than scalar operations implemented in a loop.

This note is organized as follows. Section 2: (i) presents notation and background on the standard data re-sampling approach for the non-parametric bootstrap; (ii) describes the multinomial sampling formulation of the bootstrap; and (iii) explains how to vectorize the bootstrap calculations. Section 3 reports a comparison of computation times (in R) required by the vectorized and standard approaches, when bootstrapping Pearson's sample correlation coefficient in real and simulated data sets. Finally, Section 4 presents some final remarks, and point out that the Bayesian bootstrap computations can also be easily vectorized.

\section{The vectorized non-parametric bootstrap}

Let $X$ represent a random variable distributed according to an unknown probability distribution $F$, and let $\bfx = (x_1, \ldots, x_N)^t$ be an observed random sample from $F$. The goal of the bootstrap is to estimate a parameter of interest, $\theta$, based on a statistic $\hat{\theta} = s(\bfx)$.

Let $\hat{F}$ represent the empirical distribution of the observed data, $\bfx$, assigning probability $1/N$ to each observed value $x_i$, $i = 1, \ldots, N$. A bootstrap sample, $\bfx^\ast = (x_1^\ast, \ldots, x_N^\ast)^t$, corresponds to a random sample of size $N$ draw from $\hat{F}$. Operationally, sampling from $\hat{F}$ is equivalent to sampling $N$ data points with replacement from the population of $N$ objects $(x_1, \ldots, x_N)$. The star notation indicates that $\bfx^\ast$ is not the actual data set $\bfx$ but rather a re-sampled version of $\bfx$. The sampling distribution of estimator $\hat{\theta}$ is then estimated from $B$ bootstrap replications of $\hat{\theta}^\ast = s(\bfx^\ast)$.

Now, consider the estimation of the first moment of the unknown probability distribution $F$,
\begin{equation}
\theta = E_F(X)~,
\end{equation}
on the basis of the observed data $\bfx$. If no further information (other than the observed sample $\bfx$) is available about $F$, then it follows that the best estimator of $\theta$ is the plug-in estimate (see page 36 of \cite{boot1993}),
\begin{equation}
\hat{\theta}  \, = \, E_{\hat{F}}(X) \, = \, \sum_{i = 1}^{N} x_i \, P_{\hat{F}}(X = x_i) \, = \, \sum_{i = 1}^{N} x_i \frac{1}{N} \, = \, s(\bfx)~,
\end{equation}
and the bootstrap distribution of $\hat{\theta}$ is generated from $B$ bootstrap replications of $\hat{\theta}^\ast = s(\bfx^{\ast}) = N^{-1} \sum_{i = 1}^{N} x_i^{\ast}$. Algorithm \ref{alg:usualboot} summarizes the approach.

\begin{algorithm}
\caption{Non-parametric bootstrap for $E_{\hat{F}}(X)$ via data re-sampling}\label{alg:usualboot}
\begin{algorithmic}[1]
\BState \emph{For $b = 1, \ldots, B$:}
\begin{itemize}
\item Draw a bootstrap sample $\bfx^\ast = (x_1^\ast, \ldots, x_N^\ast)^t$ from the empirical distribution of the observed data, that is, sample $N$ data points with replacement from the population of $N$ objects $\bfx = (x_1, \ldots, x_N)^t$.
\item Compute the bootstrap replication $\hat{\theta}^\ast = N^{-1} \sum_{i = 1}^{N} x_i^{\ast}$.
\end{itemize}
\end{algorithmic}
\end{algorithm}

Alternatively, let $n_i^\ast$ represent the number of times that data point $x_i$ appears in the bootstrap sample $\bfx^\ast$, and $w_i^\ast = n_i^\ast/N$. Then, the category counts, $\bfn^\ast = (n_1^\ast, \ldots, n_N^\ast)^t$, of the bootstrap sample $\bfx^\ast$ are distributed according to the multinomial distribution,
\begin{equation}
\bfn^\ast \, = \, N \, \hat{\bfw}^\ast \; \sim \; \mbox{Multinomial}\left( N \, , \, N^{-1} \, \bfone_{N} \right)~,
\label{eq:boot.mult}
\end{equation}
where the vector $\bfone_{N} = (1, \ldots, 1)^t$ has length $N$.

Now, since
\begin{equation}
\sum_{i = 1}^{N} x_i^{\ast} \, = \, \sum_{i = 1}^{N} n_i^\ast \, x_i~,
\label{eq:main.connection}
\end{equation}
it follows that the bootstrap replication of the first sample moment of the observed data can we re-expressed, in terms of the bootstrap weights $\bfw^\ast$ as,
\begin{equation}
\hat{\theta}^\ast \, = \, \frac{1}{N} \, \sum_{i = 1}^{N} x_i^{\ast} \, = \, \sum_{i = 1}^{N} w_i^\ast \, x_i~,
\end{equation}
so that we can generate bootstrap replicates using Algorithm \ref{alg:multiboot}.

\begin{algorithm}
\caption{Non-parametric bootstrap for $E_{\hat{F}}(X)$ via multinomial sampling}\label{alg:multiboot}
\begin{algorithmic}[1]
\BState \emph{For $b = 1, \ldots, B$:}
\begin{itemize}
\item Draw a bootstrap count vector $\bfn^\ast = (n_1^\ast, \ldots, n_N^\ast)^t \, \sim \, \mbox{Multinomial}\left( N \, , \, N^{-1} \, \bfone_{N} \right)$.
\item Compute the bootstrap weights $\bfw^\ast = (n_1^\ast/N, \ldots, n_N^\ast/N)^t$.
\item Compute the bootstrap replication $\hat{\theta}^\ast = \sum_{i = 1}^{N} w_i^{\ast} \, x_i$.
\end{itemize}
\end{algorithmic}
\end{algorithm}

The main advantage of this multinomial sampling formulation of the non-parametric bootstrap is that it allows the vectorization of the computation. Explicitly, Algorithm \ref{alg:multiboot} can be vectorized as follows:
\begin{enumerate}
\item Draw $B$ bootstrap count vectors, $\bfn^\ast$, from (\ref{eq:boot.mult}), using a single call of a multinomial random vector generator (e.g., \texttt{rmultinom} in R).
\item Divide the sampled bootstrap count vectors by $N$ in order to obtain a $N \times B$ bootstrap weights matrix, $\bfW^\ast$.
\item Generate the entire vector of bootstrap replications,
\begin{equation}
\hat{\bftht}^\ast \, = \, \bfx^t \, \bfW^\ast
\end{equation}
in a single computation.
\end{enumerate}

It is clear from equation (\ref{eq:main.connection}) that this multinomial sampling formulation is available for statistics based on any arbitrary sample moment (that is, statistics defined as functions of arbitrary sample moments). For instance, the sample correlation between data vectors $\bfz = (z_1, \ldots, z_N)^t$ and $\bfy = (y_1, \ldots, y_N)^t$, is a function of the sample moments,

\begin{equation}
\frac{1}{N} \sum_{i=1}^{N} z_i~, \hspace{0.3cm} \frac{1}{N} \sum_{i=1}^{N} y_i~, \hspace{0.3cm} \frac{1}{N} \sum_{i=1}^{N} z_i^2~, \hspace{0.3cm} \frac{1}{N} \sum_{i=1}^{N} y_i^2~, \hspace{0.3cm} \frac{1}{N} \sum_{i=1}^{N} z_i \, y_i~,
\end{equation}
and the bootstrap replication,
\begin{equation}
\hat{\theta}^\ast \, = \ \frac{N \sum_{i} z_i^\ast \, y_i^\ast \, - \, (\sum_{i} z_i^\ast) \, (\sum_{i} y_i^\ast)}{\sqrt{\big[N \sum_{i} z_i^{\ast2} - (\sum_{i} z_i^\ast)^2\big] \, \big[N \sum_{i} y_i^{\ast 2} - (\sum_{i} y_i^\ast)^2\big]}}~,
\end{equation}
can be re-expressed in terms of bootstrap weights as,
\begin{equation}
\hat{\theta}^\ast \, = \, \dfrac{\sum_i w_i^\ast \, z_i \, y_i \, - \, (\sum_{i} w_i^\ast \, z_i) (\sum_{i} w_i^\ast \, y_i)}{\sqrt{\big[\sum_i w_i^\ast \, z_i^2 - (\sum_{i} w_i^\ast \, z_i)^2\big] \, \big[\sum_i w_i^\ast \, y_i^2 - (\sum_{i} w_i^\ast \, y_i)^2\big]}}~,
\end{equation}
and in vectorized form as,
\begin{equation}
\hat{\bftht}^\ast \, = \, \dfrac{(\bfz \bullet \bfy)^t \bfW^\ast \, - \, (\bfz^t \bfW^\ast) \bullet (\bfy^t \bfW^\ast)}{\sqrt{\big[(\bfz^2)^t \bfW^\ast - (\bfz \bfW^\ast)^2 \big] \bullet \big[(\bfy^2)^t \bfW^\ast - (\bfy^t \bfW^\ast)^2 \big]}}~,
\label{eq:vec.cor}
\end{equation}
where the $\bullet$ operator represents the Hadamard product of two vectors (that is, the element-wise product of the vectors entries), and the square and square root operations in the denominator of (\ref{eq:vec.cor}) are also performed entry-wise.

\section{Illustrations}

In this section, we illustrate the gain in computational efficiency achieved by the vectorized multinomial sampling bootstrap, relative to two versions the standard data re-sampling approach: (i) a strait forward version based on a \texttt{for} loop; and (ii) a more sophisticated version, implemented in the \texttt{bootstrap} R package\cite{bootstrapRpackage}, where the \texttt{for} loop is replaced by a call to the \texttt{apply} R function. In the following, we refer to these two versions as ``loop" and ``apply", respectively.

\begin{figure}[!h]
\begin{center}
\includegraphics[angle=270, scale = 0.65, clip]{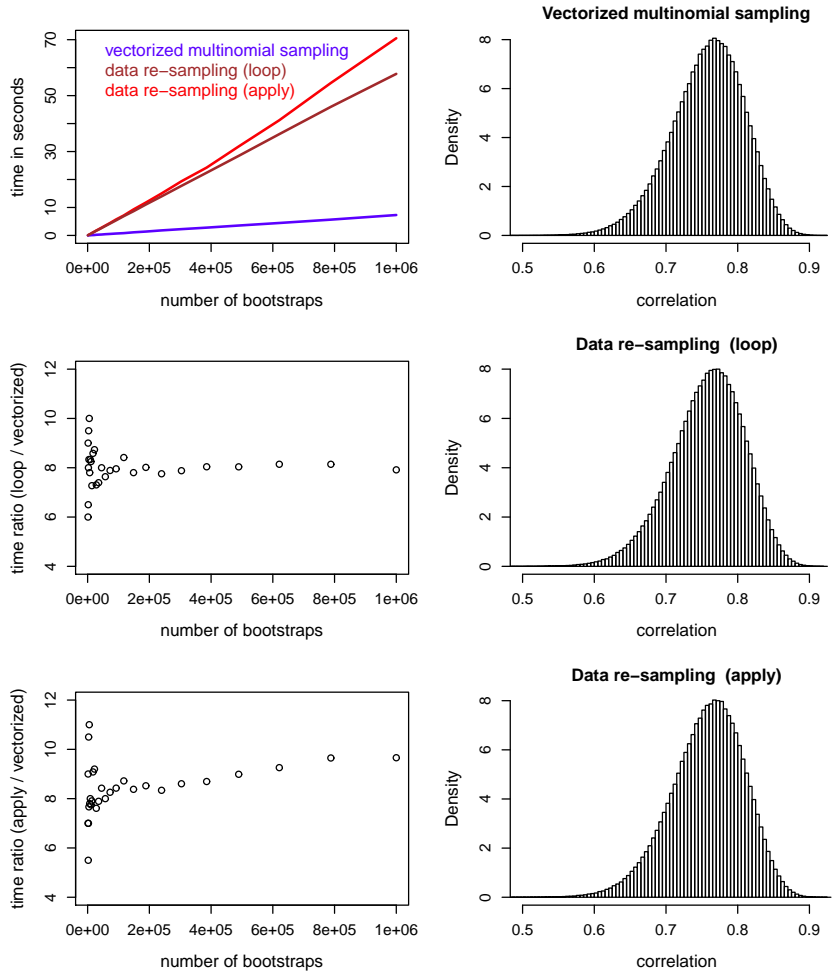}
\caption{Comparison between the data re-sampling approaches and the vectorized multinomial sampling bootstrap, in the American law school data. The top left panel compares the approach's time expenditures. The center and bottom left panels the computation time ratio of the data re-sampling versus the vectorized approaches. The right panels present the $\hat{\theta}^\ast = \hat{\mbox{cor}}(\mbox{LSAT}^\ast, \mbox{GPA}^\ast)$ distributions generated with $B = 1,000,000$.}
\label{fig:example1}
\end{center}
\end{figure}

We bootstrapped Pearson's sample correlation coefficient using the American law school data (page 21 of \cite{boot1993}) provided in the \texttt{law82} data object of the \texttt{bootstrap} R package. The data is composed of two measurements (class mean score on a national law test, LSAT, and class mean undergraduate grade point average, GPA) on the entering class of 1973 for $N = 82$ American law schools. Figure \ref{fig:example1} presents the results.

The top left panel of Figure \ref{fig:example1} shows the time (in seconds) required to generate $B$ bootstrap replications of $\hat{\theta}^\ast = \hat{\mbox{cor}}(\mbox{LSAT}^\ast, \mbox{GPA}^\ast)$, for $B$ varying from 1,000 to 1,000,000. The red, brown, and blue lines show, respectively, the computation time required by the ``apply", ``loop", and the vectorized multinomial sampling approaches. The center and bottom left panels show the computation time ratio of the data re-sampling approach versus the vectorized approach as a function of $B$. The plots clearly show that the vectorized bootstrap was considerably faster than the data re-sampling implementations for all $B$ tested. The right panels show the $\hat{\theta}^\ast$ distributions for the three bootstrap approaches based on $B = 1,000,000$.

\begin{figure}[!h]
\begin{center}
\includegraphics[angle=270, scale = 0.65, clip]{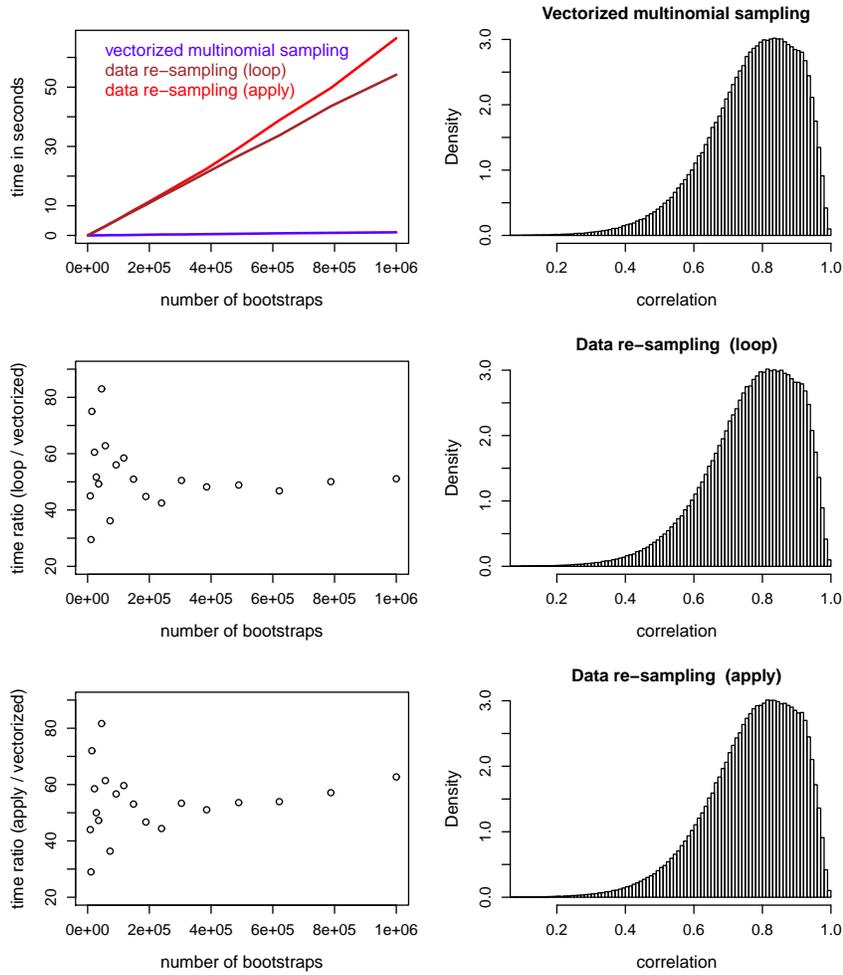}
\caption{Comparison between the data re-sampling approaches and the vectorized multinomial sampling bootstrap, in a subset of the American law school data. The top left panel compares the approach's time expenditures. The center and bottom left panels the computation time ratio of the data re-sampling versus the vectorized approaches. The right panels present the $\hat{\theta}^\ast = \hat{\mbox{cor}}(\mbox{LSAT}^\ast, \mbox{GPA}^\ast)$ distributions generated with $B = 1,000,000$.}
\label{fig:example2}
\end{center}
\end{figure}

Figure \ref{fig:example2} presents analogous comparisons, but now focusing on a subset of $N = 15$ samples from the American law school data (page 19 of \cite{boot1993}), provided in the \texttt{law} data object of the \texttt{bootstrap} R package. This time, the vectorized implementation was remarkably faster than the data re-sampling versions. The center and bottom left panels of Figure \ref{fig:example2} show that the vectorized implementation was roughly 50 times faster than the re-sampling versions, whereas in the previous example it was about 8 times faster (center and bottom left panels of Figure \ref{fig:example1}).

\begin{figure}[!h]
\begin{center}
\includegraphics[angle=270, scale = 0.65, clip]{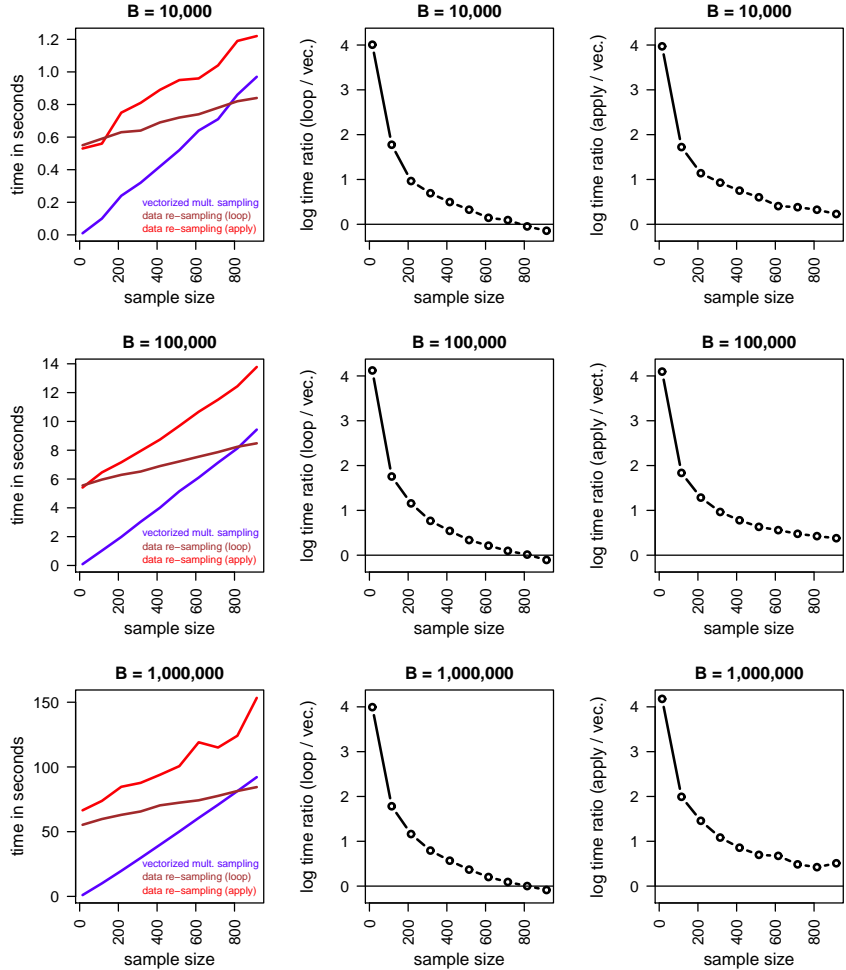}
\caption{Comparison of the bootstrap implementations, when bootstrapping Pearson's sample correlation, $\hat{\theta}^\ast = \hat{\mbox{cor}}(\bfx_1^\ast, \bfx_2^\ast)$, for sample sizes varying from 15 to 915. The data was simulated according to $x_{1i} = \epsilon_{1i}$, $x_{2i} = x_{1i} + \epsilon_{2i}$, $\epsilon_{ji} \sim \mbox{N}(0, 1)$, $j = 1, 2$. Results based on 10,000 (top panels), 100,000 (center panels), and 1,000,000 (bottom panels) bootstrap replications. The center and right panels show the computation time ratios (in log scale) comparing the ``loop" vs vectorized and the ``apply" vs vectorized implementations, respectively.}
\label{fig:example3}
\end{center}
\end{figure}

The performance difference observed in these two examples suggests that the gain in speed achieved by the vectorized implementation decreases as a function of the sample size. In order to confirm this observation, we used simulated data to compare the bootstrap implementations across a grid of 10 distinct sample sizes (varying from 15 to 915) for $B$ equal to 10,000, 100,000, and 1,000,000. Figure \ref{fig:example3} reports the results.

The left panels of Figure \ref{fig:example3} show computation times as a function of increasing sample sizes. In all cases tested the vectorized implementation (blue line) outperformed the ``apply" version (red line), whereas the ``loop" version (brown line) outperformed the vectorized implementation for large sample sizes (but was considerably slower for small and moderate sample sizes). The central and right panels show the computation time ratios (in log scale) comparing, respectively, the ``loop" vs vectorized and the ``apply" vs vectorized implementations. The horizontal line is set at zero and represents the threshold below which the data re-sampling approach outperforms the vectorized implementation. Note that log time ratios equal to 4, 3, 2, 1, and 0.5, correspond to speed gains of 54.60, 20.09, 7.39, 2.72, and 1.65, respectively.

All the timings in this section were measured on an Intel Core i7-3610QM (2.3 GHz), 24 Gb RAM, Windows 7 Enterprize (64-bit) platform.

\section{Discussion}

In this note we showed how the multinomial sampling formulation of the non-parametric bootstrap can be easily implemented in vectorized form. We illustrate the gain in computational speed (in the R programming language) using real and simulated data sets.

Our examples provide several interesting insights. First, the re-sampling implementation based on the \texttt{for} loop was generally faster than the implementation provided by the \texttt{bootstrap} R package, which employs the \texttt{apply} in place of the \texttt{for} loop (compare the red and brown curves on the top left panel of Figures \ref{fig:example1} and \ref{fig:example2}, and on the left panels of Figure \ref{fig:example3}). This result illustrates the fact that \texttt{apply} is not always faster than a \texttt{for} loop (see \cite{ligges2008} for further discussion and examples). Second, the vectorized implementation outperformed the data re-sampling implementation provided in the \texttt{bootstrap} R package in all cases tested (left panels in Figures \ref{fig:example1} and \ref{fig:example2}, and right panels in Figure \ref{fig:example3}). Third, the gain in speed achieved by the vectorized implementation decreases as a function of the sample size (Figure \ref{fig:example3}). This decrease is likely due to the increase in memory requirements for generating and performing operations in the larger bootstrap weight matrices associated with the larger sample sizes. We point out, however, that even though optimized BLAS (Basic Linear Algebra Subprograms) libraries could potentially increase the execution speed of our vectorized operations in large matrices/vectors \cite{ligges2008}, our examples still show remarkable/considerable gains for small/moderate sample sizes even without using any optimized BLAS library (as illustrated by Figures \ref{fig:example2} and \ref{fig:example1}).

For the sake of clarity, the exposition in Section 2 focused on statistics based on sample moments of the observed data, $\bfx$. We point out, however, that the multinomial sampling formulation of the bootstrap is available for any statistic satisfying the more general relation,
\begin{equation}
\sum_{i = 1}^{N} \, f(x_i^\ast) \, = \, \sum_{i = 1}^{N} \, n_i^\ast \, f(x_i)~,
\label{eq:general}
\end{equation}
where the $k$th sample moment represents the particular case, $f(x_i) = x_i^k/N$. Clearly, the left hand side of equation (\ref{eq:general}) still represents a bootstrap replication of the first sample moment of the transformed variable $u_i^\ast = N f(x_i^\ast)$.

The multinomial sampling formulation of the non-parametric bootstrap is not new. It is actually a key piece in the demonstration of the connection between the non-parametric bootstrap and Bayesian inference, described in \cite{rubin1981} and in Section 8.4 of \cite{eosl2001}, where the non-parametric bootstrap is shown to closely approximate the posterior distribution of the quantity of interest generated by the Bayesian bootstrap \cite{rubin1981}. This close connection, also implies that the Bayesian bootstrap can be easily implemented in vectorized form. As a matter of fact, instead of generating the bootstrap weights from bootstrap count vectors sampled from a multinomial distribution, the Bayesian bootstrap samples the weights directly from a $\mbox{Dirichlet}(\bfone_N)$ distribution. We point out, however, that, to the best of our knowlege, the multinomial sampling formulation has not been explored before for vectorizing bootstrap computations.

\end{document}